\documentclass[twocolumn,floatfix,prl,showpacs]{revtex4}

\usepackage[dvips]{graphicx}
\usepackage{dcolumn}
\usepackage{bm}
\usepackage{amssymb}
\usepackage[usenames]{color}

\begin{document}

\title{Non-local composite spin-lattice polarons in high temperature superconductors}
\author{G. De Filippis$^1$, V. Cataudella$^1$, A. S. Mishchenko$^{2,3}$, and 
N. Nagaosa$^{2,4}$}
\affiliation{$^1$Coherentia-CNR-INFM and Dip. di Scienze Fisiche - Universit\`{a} di Napoli
Federico II - I-80126 Napoli, Italy\\
$^2$CREST, Japan Science and Technology Agency (JST) - AIST - 1-1-1 -
Higashi - Tsukuba 305-8562 - Japan\\
$^3$RRC ``Kurchatov Institute'' - 123182 - Moscow - Russia\\
$^4$CREST, Department of Applied Physics, The University of Tokyo, 7-3-1 Hongo, 
Bunkyo-ku, Tokyo 113, Japan}
\date{\today}

\begin{abstract}
The non-local nature of the polaron formation in t-t$'$-t$''$-J
model is studied in large lattices up to 64 sites
by developing a new numerical method. 
We show that the effect of longer-range hoppings t$'$ and t$''$ is a 
large anisotropy of the electron-phonon 
interaction (EPI) leading to a completely different influence 
of EPI on the nodal and antinodal points in agreement with the
experiments. Furthermore, nonlocal EPI preserves polaron's quantum motion, 
which destroys the antiferromagnetic order effectively, 
even at strong coupling regime, 
although the quasi-particle weight in angle-resolved-photoemission 
spectroscopy is strongly suppressed.
\end{abstract}

\pacs{71.10.Fd, 71.38.-k, 02.70.Ss, 75.50.Ee}
\maketitle

It is a matter of long debates whether electron-phonon interaction 
(EPI) plays an essential role in the formation of the puzzling
properties of high temperature superconductors \cite{Shen_03}. 
Early theoretical studies of the angle resolved photoemission 
spectroscopy (ARPES) of undoped cuprates were based on the 
t-t$'$-t$''$-J model, where doped holes into 
an antiferromagnetic (AFM) material are characterized by 
exchange constant J 
and hoppings up to 3rd near neighbors(NN) 
with amplitudes t, t$'$ and t$''$. This approach  
successfully described the dispersion of the 
experimentally observed peak \cite{xiang} and 
appeared to exclude strong EPI. 
However, there was a strong contradiction between the theoretical and 
experimental lineshapes: in contrast with the very broad 
peak in experiments \cite{Shen_03} the theory predicted a sharp peak 
\cite{DagMan,liu,tJ01}.    
This contradiction was resolved under assumption that the undoped parent 
compounds are in the strong coupling regime (SCR) of the EPI. 
In the simplest case of t-J-Holstein model, where a hole interacts 
with dispersionless optical phonons through on-site local coupling, 
it was shown that at SCR the spectral weight of the
quasiparticle is almost completely transferred to the broad Franck-Condon 
shake-off peak which inherits the dispersion of the hole when it does not interact 
with phonons \cite{andrei}. 
Furthermore, the inheritance of the dispersion 
of the noninteracting quasiparticle by the Franck-Condon peak was   
shown to be a property of a broad class of models with EPI \cite{RoGu2005}. 
Recently, the relevance of the t-J-Holstein model for   
cuprates has been rather convincingly questioned\cite{horsch}.
It has been shown that SCR leads to a picture of a localized static hole 
with 4 broken bonds about it.
In this case the percolative model predicts that AFM order survives up 
to critical concentration $x \approx 0.5$ in contradiction with
the experimental value $x \approx 0.02-0.04$. 
Another problem of the t-J-Holstein model is the very large effective 
mass in the whole SCR \cite{andrei} contradicting 
transport and optical properties of lightly doped cuprates \cite{Lee05}.

The above arguments, in favor or against the SCR in undoped 
cuprates, suggest that the answer must be found in a class of 
more realistic t-t$'$-t$''$-J models where EPI, as it is in 
cuprates \cite{Ishihara}, is nonlocal. 
Besides of the particular relevance of the 
above models in the physics of high temperature superconductivity,
there is a general inquiry asking whether long-range interaction 
can introduce profound qualitative difference compared with 
t-J-Holstein model. 
However, study of the t-t$'$-t$''$-J model with nonlocal EPI by 
exact diagonalization (ED) and Diagrammatic Monte Carlo (DMC) 
methods is hindered by their limitations. 
ED method has been restricted to only small two-dimensional systems 
\cite{horsch, fehske} 
and the nonlocal effects have not been studied so far. 
On the other hand the study of the incoherent motion of the hole   
requires transformation of the DMC method \cite{andrei} into the direct 
space where DMC encounters the sign problem \cite{tJ01}.    

In this Letter we apply recently introduced coherent state 
basis approach 
\cite{noi} to the t-t$'$-t$''$-J model with nonlocal EPI, 
and develop a Coherent States Lanczos (CSL) method which
is able to provide a reliable description of the ground state properties 
of two-dimensional systems up to $64$ sites. 
We validate it by successfully comparing the results with exact DMC data.   
We find that in the t-t$'$-t$''$-J model with local Holstein EPI, 
in contrast with t-J-Holstein model, the influence of the EPI on the 
properties of nodal ${\bf k} = (\pi /2 , \pi /2)$ and antinodal 
${\bf k} = (\pi, 0)$ points is significantly different
introducing a considerable anisotropy driven 
by long-range hole hoppings: the antinodal states are in the SCR with 
nearly zero quasiparticle weight whereas the quasiparticles in the 
nodal point are only weakly dressed by phonons in a very broad 
range of EPI. 
We also show that nonlocal EPI qualitatively changes the basic features 
of the composite spin-lattice polarons.
First, according to general result \cite{Kornilovitch}, effective mass in 
the SCR of nonlocal EPI is considerably lighter than that of a model with 
local EPI.  
Second, the motion of the hole over the NN is not strongly 
suppressed  by nonlocal EPI and, thus, the tendency toward robust AFM 
order up to high doping levels, found in t-J-Holstein model, does not 
survive in the more realistic  t-t$'$-t$''$-J model with nonlocal EPI. 

The minimal Hamiltonian for t-t$'$-t$''$-J model with extended EPI is a sum of 
t-t$'$-t$''$-J Hamiltonian $H_{tt^{'}t^{''}J}$
\cite{xiang,DagMan,Kane89}, EPI Hamiltonian $H_{h-ph}$, 
and Hamiltonian of dispersionless phonons 
$H_{ph}=\omega_0 \sum_i c_i^{\dagger }c_i$ 
($c_i^{\dagger }$ is the creation operator of a phonon at site i with 
frequency $\omega_0$).
In the spin wave approximation the t-t$'$-t$''$-J Hamiltonian reads
\begin{eqnarray}
&&H_{tt^{'}t^{''}J}=\sum_{\vec{q}} \left [\Omega_{\vec{q}} \left( a_{\vec{q}}^{\dagger }
a_{\vec{q}}+ b_{\vec{q}}^{\dagger }b_{\vec{q}} \right) + \epsilon_{\vec{q}}
\left( h_{\vec{q}}^{\dagger }h_{\vec{q}}+ f_{\vec{q}}^{\dagger }f_{\vec{q}} \right)
\right]+ \nonumber \\
&& \sum_{\vec{q}} \sum_{\vec{\delta}} \left( \sum_{i \epsilon A} 
M_{\vec{q},i}  f_i^{\dagger} h_{i+\delta} a_{\vec{q}} + 
\sum_{i \epsilon B} M_{\vec{q},i} h_i^{\dagger} f_{i+\delta}b_{\vec{q}}+h.c.  
\right) \nonumber,
\end{eqnarray}
where $A$ ($B$) is the sublattice with spin up (down) and 
$a_{\vec{q}}^{\dagger }$ ($b_{\vec{q}}^{\dagger }$) is the operator
creating magnon in the A (B) sublattice with dispersion $\Omega_{\vec{q}}$ 
\cite{DagMan,liu,tJ01} .
The operator in the direct space $f_{i}^{\dagger }$ ($h_{i}^{\dagger }$) 
creates a spinless hole on the site $i$ of the sublattice A (B) . 
The bare hole dispersion is 
$\epsilon_{\vec{q}}=4t^{'}\cos(q_x) \cos(q_y)+ 2t^{''}(\cos(2q_x)+\cos(2q_y))$
and the hole-magnon coupling is  
$ M_{\vec{q},i}=t \sqrt {2/N} \left( u_{\vec{q}}e^{i \vec{q} \cdot \vec{R}_{i}}+ 
v_{\vec{q}}e^{i \vec{q} \cdot \vec{R}_{i+\delta}} \right)$, where $u_{\vec{q}}$ and 
$v_{\vec{q}}$ are the Bogoliubov factors,  
$\vec{\delta}$ is an unitary vector connecting NN,  
and $N$ is the number of lattice sites. 
The sum over ${\vec{q}}$ is restricted inside the magnetic Brillouin zone. 
The EPI Hamiltonian 
\begin{eqnarray}
H_{h-ph} &=& \omega_0 \sum_l g(l) \sum_{i \epsilon A} f_i^{\dagger}f_i \left( c_{i+l}^{\dagger}+c_{i+l} \right)+ \nonumber \\
&& \omega_0 \sum_l g(l) \sum_{i \epsilon B} h_i^{\dagger }h_i \left( c_{i+l}^{\dagger}+c_{i+l} \right) 
\label{Hepi}
\end{eqnarray} 
in defined in terms of Holstein, $g(0)=g$, and nonlocal coupling to the 
NN lattice displacements $g(\vec{\delta})=g_1$. Below we set
$\hbar=1$ and $t=1$.

\begin{figure}
       \includegraphics[scale=0.45]{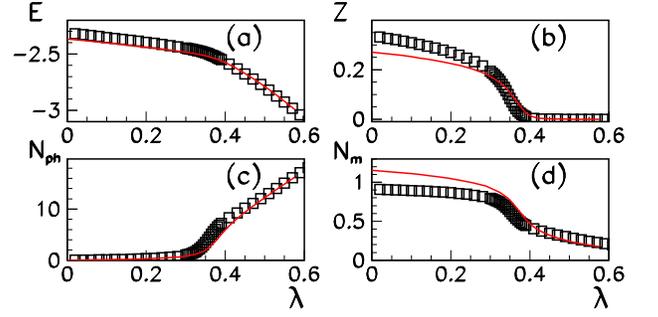}
        \caption{(Color online)
Comparison of energy (a), spectral weight (b), 
phonon (c) and magnon (d) average numbers (squares) obtained by  
CSL method at $N=64$ with those obtained by DMC method (line) for the 
t-J-Holstein model at $J/t=0.3$, $\omega_0/t=0.1$,
and ${\bf k}=(\pi/2,\pi/2)$.}
\end{figure}

In order to study the ground state of this 
model, we 
use CSL procedure based on the Lanczos recursion method that,  
starting from the hole in the quantum Neel state without excited magnons, 
$\left| 0\right\rangle^{_{(m)}}$, generates the subspace spanned by 
the basis 
\begin{eqnarray}
\left| {j},{\mu_1},...,{\mu_N},{\vec{q}_1,...,\vec{q}_{l},l} \right\rangle =
\left|h \right\rangle_j
\left [ \prod_{i} \left| \mu_i\right\rangle \right]  \left|   \vec{q}_1,....,\vec{q}_{l},l  \right\rangle \nonumber, 
\end{eqnarray}
where $\left|h \right\rangle_j$ indicates the state with the hole on the 
site $j$, and $i$ runs over the lattice sites.
The notation $\left| \vec{q}_1,....,\vec{q}_{l},l\right\rangle$ labels 
the standard Bonfim-Reiter $l$-magnon states which are sufficient 
to reproduce the results of self-consistent Born approximation (SCBA)
\cite{ramsak}.  
These states are given in terms of an ordered product of 
$a_{\vec{q}}^{\dagger }$ and $b_{\vec{q}}^{\dagger }$ operators acting 
on the magnon vacuum state
$
\left|   \vec{q}_1,....,\vec{q}_{l},l  \right\rangle=a_{\vec{q}_{l}}^{\dagger}b_{\vec{q}_{l-1}}^{\dagger}....
\left| 0\right\rangle^{_{(m)}}  
$
($\left|   \vec{q}_1,....,\vec{q}_{l},l  \right\rangle=b_{\vec{q}_{l}}^{\dagger}a_{\vec{q}_{l-1}}^{\dagger}....
\left| 0\right\rangle^{_{(m)}} \
$) 
if the hole is on $A$ ($B$) sublattice.
Within the Bonfim-Reiter basis, 
the Lanczos or any other ED procedure is capable of reproducing
the ground state properties and spectral functions within the SCBA
if the following approximations are implemented: 
1) the magnons involved into the 
string 
$a_{\vec{q}_{l}}^{\dagger}b_{\vec{q}_{l-1}}^{\dagger}....
\left| 0\right\rangle^{_{(m)}}$ are considered to be distinguishable, 
leading thus to the unity normalization of the state 
$\left|   \vec{q}_1,....,\vec{q}_{l},l  \right\rangle$; 2) 
the hole-magnon coupling term, as in the retraceable-path approximation \cite{rice},
is considered to be able to annihilate only the last excited magnon. 
For the study of the ground state properties the situation is 
considerably simplified because for $J/t \ge 0.3$ the states containing 
less than 4 magnons are sufficient to reproduce the results of SCBA 
\cite{ramsak}. The real bottleneck for 
increasing the size of the system comes from 
the phonon basis states 
$\left| \mu _{i} \right\rangle = 
\frac{\left(c^{\dagger}_{i}\right)^{\mu_{i}}}{\sqrt{\mu_{i}!}}
\left| 0\right\rangle _{i}^{_{(ph)}}$, 
where $\left| 0\right\rangle _{i}^{_{(ph)}}$ is the phonon vacuum at the
site $i$. The number of relevant phonon states quickly grows at 
intermediate values of the EPI strength restricting the 
lattice sizes to 10 sites \cite{fehske}, the maximal size we are aware of, for 
quantum phonons and to 20 sites in the fully adiabatic limit \cite{horsch}.  
\begin{figure}
        \includegraphics[scale=0.45]{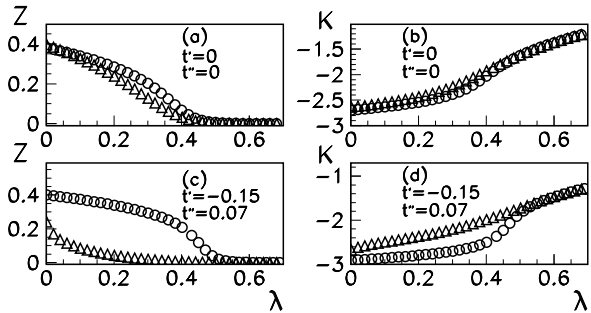}
\caption{The spectral weight (a and c) and the mean kinetic energy 
(b and d) as function of $\lambda$ for nodal (circles) and antinodal 
(triangles) points. 
The parameter values are: 
$J/t=0.4$, $\omega_0/t=0.2$, $g_1=0$, $N=64$.}
\end{figure}
To circumvent the bottleneck we switch from the basis of 
many-phonon states 
$\left| \mu _{i} \right\rangle$ to that expressed in terms of coherent states 
(CS) \cite{noi} which are nothing but the canonical transformations for 
independent oscillator model 
\begin{equation}
\left| h,i\right\rangle =e^{gh(b_{i}-b_{i}^{\dagger })}\left| 0\right\rangle
_{i}^{_{(ph)}}=e^{-\frac{g^2 h^2}{2}} \sum_{n=0}^{\infty} \frac {\left(-gh \right)^n}{\sqrt{n!}}
\left| n\right\rangle_{i} .
\label{Coherent}
\end{equation}
They are renormalized by different parameters $h$. For 
$h=0$ the CS is the bare state without EPI and for $h=1$ it is 
the exact solution of the independent oscillator model with local 
coupling $g$.
In principle, this substitution does not introduce any truncation in 
the Hilbert space since, varying $h$ in the complex plane, the local 
basis (\ref{Coherent}) is over-complete. 
Naturally, particular realization of the CSL approach  
requires a truncation. Actually only few CS are enough to 
reproduce the ground state properties of the composite spin-lattice polaron.
Comparison with exact DMC data shows that the following truncation 
scheme minimizes the computational efforts with high accuracy 
for results. 
A finite number $M_{ph}$ of CS is chosen, 
characterized by real values $h_{\alpha }$\ ($\alpha =0,...,M_{ph}-1$) 
equidistant in the range $[0,1]$, and only the phonon states 
$\prod_{i=1}^{N}\left|h_{\alpha _{i}},i\right\rangle$ with 
$\sum_{i=1}^{N} \alpha_{i} \le M_{c,ph}$ are included into the basis.
Besides, taking advantage of the fact that the hopping and EPI 
are restricted to 3rd NN and NN respectively, we restrict the states 
with nonzero parameters $h$ to 3rd NN from
the hole.   
Comparison (Fig.~1) of CSL results for t-J-Holstein model on $8 \times 8$ 
lattice with approximation-free result by DMC in the thermodynamic limit 
\cite{andrei} shows that our method at $M_{ph} = 4$ and $M_{c,ph} = 3$ 
is valid for all ranges of EPI \cite{convergence}.  
The small differences in the weak EPI regime are mainly due to finite size 
effects. 
At large dimensionless couplings $\lambda=g^2\omega_0/4t$ 
the polaron size becomes small and the agreement becomes excellent. 
Our method perfectly reproduces crossover between weak and 
strong coupling regimes at critical value $\lambda_c \approx 0.4$ where i)
the spectral weight goes to zero, indicating suppression of the coherent motion; 
ii) the lattice distortions rapidly increase; and iii) the size of
magnetic polaron reduces.

\begin{figure}
        \includegraphics[scale=0.45]{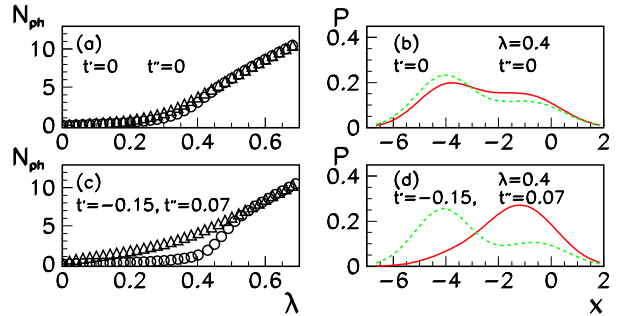}
\caption{(Color online) 
Average number of phonons (a and c) for nodal (circles) 
and antinodal (triangles) points; PPDF (b and d) 
for nodal (solid lines) and antinodal (dashed lines) points. 
The parameter values are:
$J/t=0.4$, $\omega_0/t=0.2$, $g_1=0$, $N=64$.}
\end{figure}

\begin{figure*}
\includegraphics[scale=0.45]{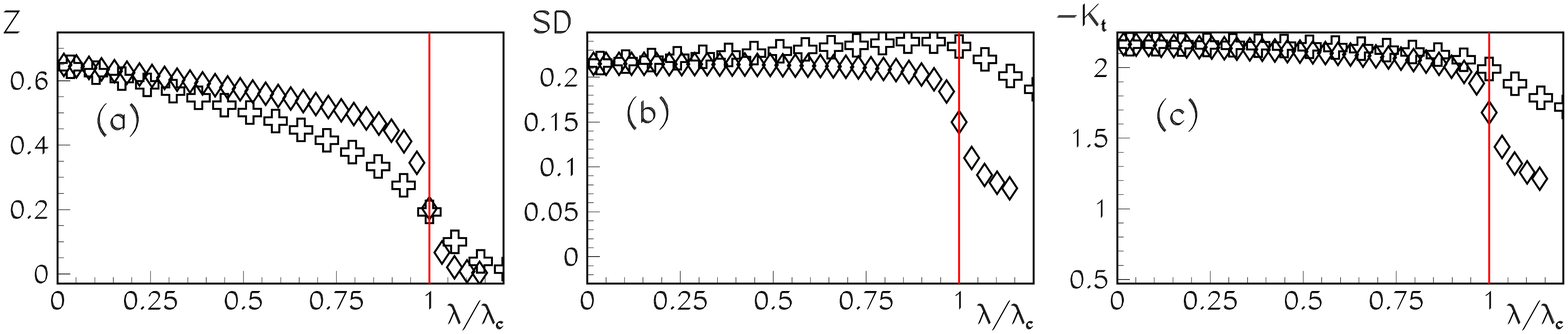}
\caption{ (Color online)
The spectral weight (a),
the spin deviation around the hole (see text) (b), 
and the absolute value of the average kinetic energy 
for magnon assisted NN hoppings, 
(c), for $g_1=0$ (diamonds) and $g_1=0.5 g$ (crosses). 
The parameter values are: $J/t=0.4$, $\omega_0/t=0.2$, $t^{'}=-0.5t$, 
$t^{''}=0.4t$, $N=16$.
Critical $\lambda_c$ is defined as that where the spectral weight 
is reduced by $70 \%$ with respect to that at $\lambda=0$. }  
\end{figure*}

First, we extend t-J-Holstein model with local EPI adding 
next NN, t$'$, and next next NN, t$''$, hoppings. 
Comparing physical properties of these models we observe that the 
longer-range hoppings introduce a strong anisotropy. 
Physical properties of the states in the nodal $(\pi /2, \pi /2)$ and 
antinodal $(\pi , 0)$ points are very similar in the  t-J-Holstein model
but considerably different in the tt$'$t$''$J-Holstein model
(Fig.~2 and 3), as it is evident comparing  
$Z$-factor (Fig.~2a and 2c), mean kinetic energy (Fig.~2b and 2d), and 
average number of phonons (Fig.~3a and 3c).  
There is a wide range of the EPI strengths from the weak to intermediate 
coupling regime where the hole is a heavily dressed polaron in the antinodal 
point but an almost free particle in the nodal one. 
Comparing spectral weights and kinetic energies of the two models we 
conclude that the longer-range hoppings reinforce coherent motion in 
the nodal point and suppress it in the antinodal one. 
To characterize the anisotropy induced by t$'$ and t$''$ we calculate 
the phonon probability distributions function (PPDF)
$P(x) = \langle \Psi_{GS}\mid x \rangle \langle x \mid \Psi_{GS} \rangle$, 
where $\Psi_{GS}$ is the wave function of the ground state
and $x$ is the lattice distortion of the site where the hole is, 
just below the crossover into the SCR (Fig.~3b and 3d). 
We find that for the t-J-Holstein model both nodal and antinodal points 
display a bimodal structure characteristic of the intermediate 
coupling regime, where the ground state of the system is a quantum 
mixture of localized and delocalized states \cite{ST02}. 
To the contrary, in the tt$'$t$''$J-Holstein model the antinodal 
point has bimodal PPDF while that at the nodal point displays 
the maximum at small lattice distortions $x$, 
characteristic  
of weak coupling regime.  
The resulting picture of the anisotropic EPI, considerably stronger 
near the antinodal point than at the nodal one, is expected to be 
qualitatively similar in the case of an underdoped system and is in line with 
experimental findings. 
The explanation of the vertical dispersion of the ARPES near the 
antinodal points requires large $Z$ (weaker coupling) along the nodal
direction and much smaller $Z$ (stronger coupling) along the antinodal ones
\cite{science}.   

To study the role of nonlocal EPI we compare the properties of the 
tt$'$t$''$J-Holstein model with that including nonlocal coupling 
to the NN lattice displacements $g_1=g/2$ (Eq.(\ref{Hepi})).
In both cases 
the decrease of the $Z$-factor (Fig.~4a) indicates the suppression 
of hole ground state contribution to ARPES. However, the increase 
of the EPI 
range slightly enhances (reduces) the suppression of 
the $Z$-factor at $\lambda<\lambda_c$ ($\lambda>\lambda_c$). 
On the contrary, when EPI is non local the 
polaron effective mass, that is a measure of the coherent motion, 
becomes larger in the weak coupling regime and smaller in the SCR. 
In the latter case, the mass for nonlocal EPI at $\lambda = 1.1 \lambda_c$ 
is one order of magnitude lighter: diagonal $m_d$ (along $k_x=k_y$)
and transverse $m_t$ (along $k_x=-k_y$) masses for local EPI
are $m_d=71.4$ and $m_t=65.7$ whereas for nonlocal EPI their values 
are $m_d=7.1$ and $m_t=7.5$, respectively, in units of the 
bare effective masses ($g=g_1=0$). 
Similar effects have been discussed in a number of different models 
\cite{Kornilovitch,fugi}.  

Figure 4b and 4c show the dependence of spin deviation 
and the contribution to the kinetic energy arising 
due to magnon assisted NN hoppings, $K_t$,   
on the dimensionless coupling constant $\lambda$. Spin deviation, 
$SD=( S_{AFM} - \langle S_{NN}\rangle  ) /  S_{AFM}$, measures 
how much the neighboring spin to the hole, $S_{NN}$, deviates   
from its value in an ideal antiferromagnet ($S_{AFM}$).
Absolute value of $K_t$ and $SD$ are the measure of 
the intensity of the NN hopping and include both coherent and 
incoherent contributions. 
The decrease of the two above quantities, (Fig.~4b and 4c),  
signals on the suppression of the motion over NN as expected in presence 
of hole-phonon interaction. However, 
by increasing the EPI range,   
the motion of the hole on NN
is suppressed by the nonlocal EPI much less effectively than 
it is in the case of the Holstein coupling. This effect is one 
of the main results of this work. It can be easily interpreted 
in the SCR where the notion of the 
adiabatic potential makes sense.
For the Holstein local coupling the self-consistent adiabatic potential 
for the hole is a deep $\delta$-function, preventing, thus, the hole from 
the motion over the NN. 
To the contrary, the motion of the hole in case of a long-range EPI
is not restricted to a single site, preserving, thus, the motion 
over the NN even at SCR. This allows an effective reduction of AFM order 
also in presence of strong hole-phonon interactions. 

In conclusion, we demonstrated the important role played by the long range 
interactions in the problem of hole-phonon coupling for the
physics of high temperature superconductors. 
Long-range electron hoppings lead to a strong anisotropy of the EPI 
which is possibly observed in the angle resolved spectroscopy of cuprates.   
The most important results are that the non-local EPI makes polarons lighter
and occurs to be not effective in the suppression of the motion of the hole 
over the NN making, thus, the models with strong nonlocal EPI 
justified for description of weakly doped high temperature superconductors.   

NN acknowledges the financial support from the 
Grant-in-Aids under the Grant numbers 15104006, 16076205, and 
17105002, and NAREGI Nanoscience Project from the Ministry of 
Education, Culture, Sports, Science, and Technology, Japan and 
ASM acknowledges support of RFBR 07-0200067-a. Two of us (GDF and VC) 
acknowledge C. Castellani, M. Grilli, M. Capone and C.A. Perroni for useful discussions.


\begin{thebibliography}{99}
\bibitem{convergence} Checking convergence we found that 
$M_{ph} = 5$ and $M_{c,ph} = 6$ is enough for all data shown 
in this paper.
\end{thebibliography}
\end{document}